\documentstyle[12pt]{article}
\newcommand{\be}{\begin{equation}}
\newcommand{\ee}{\end{equation}}
\newcommand{\bea}{\begin{eqnarray}}
\newcommand{\eea}{\end{eqnarray}}
\newcommand{\bd}{\begin{displaymath}}
\newcommand{\ed}{\end{displaymath}}
\newcommand{\bi}{\begin{itemize}}
\newcommand{\ei}{\end{itemize}}
\newcommand{\bc}{\begin{center}}
\newcommand{\ec}{\end{center}}
\newcommand{\bfl}{\begin{flushleft}}
\newcommand{\efl}{\end{flushleft}}
\newcommand{\bfr}{\begin{flushright}}
\newcommand{\efr}{\end{flushright}}

\def\6{\partial}  \def\b{\beta}

  \def\t{\tau}

 \def\G{\Gamma} 
 \def\L{\Lambda} \def\S{\Sigma}
  \def\O{\Omega}

\def\={\!\!\!&=&\!\!\!}
\def\+{\!\!\!&&\!\!\!+~}
\def\-{\!\!\!&&\!\!\!-~}

\renewcommand{\AA}{{\cal A}}

\newcommand{\CC}{{\cal C}}
\newcommand{\DD}{{\cal D}}

\newcommand{\II}{{\cal I}}

\newcommand{\RR}{{\cal R}}
\newcommand{\SS}{{\cal S}}

\newcommand{\VV}{{\cal V}}

\begin{document}

\title{Topological Strings and $D$-Branes \footnote{Prepared for the {\em Second International Conference on
Fundamental Interactions}, June 6-12, 2004, Pedra Azul, Espirito Santo, Brazil}}
\author{Ion Vasile Vancea}
\maketitle
\bc
{\small \em Departamento de F\'{\i}sica Matem\'{a}tica,\\
Faculdade de Filosofia, Ci\^{e}ncias e Letras de Ribeir\~{a}o Preto,
Universidade de S\~{a}o Paulo,\\
Av. Bandeirantes 3900, 14040-901, Ribeir\~{a}o Preto - SP, Brasil}  
\ec

\begin{abstract}
In this talk we give a brief review of the algebraic structure behind the open and closed topological strings and
$D$-branes and emphasize the role of tensor category and the Frobenius algebra. Also, we speculate on the possibility of
generalizing the topological strings and the $D$-branes through the subfactor theory.
\end{abstract}

\newpage

\section{Introduction}

The $D$-branes, discovered by J. Polchinski in 1995 \cite{jp}, are extended objects in string theory characterized by 
physical properties such as tension, charge and supersymmetry and by certain geometric properties which depend on the
chosen string theory. In some spacetime backgrounds, as the flat spacetime, the geometry of $D$-branes  
can be determined postulating that the $D$-branes are those submanifolds on which the open strings can end. From that,
one infers that in the low energy limit of string theory the $D$-branes are solutions of supergravity consistent with
$T$-duality and supporting supersymmetry in various degrees. The description of $D$-branes as submanifolds of spacetime
 is valid for 
other string
theories, e. g. described by certain $\sigma$-models. Also, a microscopic description of $D$-branes as coherent 
states of 
infinite norm from the Fock space 
of closed strings can be given in the perturbative limit of string theory \cite{pv1,pv2}. However, the vacuum of string theory is 
highly degenerate. Consequently, an infinite number of theories on spacetime manifolds of various dimensions from eleven to four can be
produced. Therefore, it is desirable to have a general definition of $D$-branes which could be used in any string
theory and any background. Also, since different (dual) string theories are valid in various regions of the 
moduli space which
should be entirely covered by string field theory, one would like to have this general definition lifted to string 
field theory. 

Formulating a general definition of $D$-branes represents an open problem and an active line of research.
Progress has been made in this direction mainly in the context of Calabi-Yau backgrounds which represent
a viable alternative for compactification  down to a spacetime with four dimensions and extensions of Standard Model
living in it. In this endeavor, new mathematical tools have been
employed and results from topological field theory have been used (see, e. g. \cite{psa,mrd1,mrd2}.) 

The aim of this
talk is to give a brief review of some basic concepts which have been useful so far in studying the topological
strings and $D$-branes. Also, we are going to speculate on
the possible generalization of these objects  by using the subfactor theory applied to the 
Frobenius algebras of the topological strings. The organization of this work is as follows. In Section 2 we are going to review the relationship
between the string theory, the topological quantum field theory (TQFT) and $D$-branes and emphasize the 
algebraic structure behind
it. In Section 3 we speculate on the 
extension of the notions topological strings and $D$-branes through the subfactor theory and suggest that this extension,
if possible, could be interpreted as a (topological) string field theory of open and closed strings. We  
justify the introduction of the mathematical structures in an intuitive manner 
and refer the reader for technical details and more information on the topics to \cite{psa,mrd1,mrd2,gm,gs,fq}. 
Also, we assume that the reader is conversant with basic category theory and abstract algebra as well as with some
well established results from string theory and $D$-branes.

\section{TQFT, Topological Strings and D-Branes}

In this section, we present the mathematical structure of TQFT and topological strings and we give
the relationship between TQFT and the Frobenius algebra. The basic references used for this section
are \cite{gs,fq,bb}.

\subsection{Categories and TQFT}

The scattering processes in string theory can be derived, in principle, by borrowing the intuitive graphical
concepts from quantum field theory. This amounts to calculate the partition function $Z$ as the Polyakov sum over 
all "loops"
in the perturbation theory
\be
Z \sim \sum_{\mbox{ \small topologies}}\int~ \DD \phi_i ~ e^{S(\phi_i)},
\label{pf}
\ee
where $\phi_i$ stand for all string degrees of freedom, bosonic as well as fermionic. The "loops" that should be 
considered in string theory are $2d$ Euclidean surfaces classified by their genus. In the case of closed strings,
all $2d$ surfaces can be obtained by gluing together three fundamental surfaces: the Hartle-Hawking
tadpole (the disk $B^2$), the propagator (the cylinder) and the three-vertex (pants diagram) \cite{gs}.
The topology changing element is the three vertex since by gluing $2n$ vertices on two branches one
can construct $n$-tori. In flat spacetime background, the modular invariance between the open string theory at 
one-loop and the closed string theory at tree level (open-closed string interpretation of the cylinder diagram)
and the $T$-duality allow to pass from the open string sector to the closed string sector. In particular, 
the $D$-branes which lies on the boundary of the cylinder in the open string theory continue to lie on the
boundary in the closed string theory. However, the role of the Dirichlet and Neumann boundary conditions in the two
sectors is interchanged \cite{jp}. This construction can be used as a guide to define the $D$-branes
in a different class of string models, the topological strings. The basic requirements are:
i) the 
string partition function $Z$ be computed as the Polyakov sum over all $2d$ homeomorphic invariant
Euclidean world-sheet surfaces of the closed string;
ii) the $D$-branes be associated to the boundaries of these surfaces;
iii) to the $D$-branes are associated spaces of fields from the open string spectra. 
In order to make prediction in the topological theory, one has to give the computation rules. A simple 
inspection of the way in which the fundamental surfaces are joined to give other surfaces suggests that these rules
be algebraic. Also, they should act on the spaces of fields on the boundaries. Since the world-sheets connect various
spaces of fields, they can be viewed as mappings among these spaces. From these remarks, one recognizes that a
suitable mathematical framework to describe $D$-branes in topological string theory is that of categories. Another way 
to see that is from the observation that the  TQFT's can be formulated in terms of modular tensor categories
\cite{gs}. 

In order to make transparent the mathematical structure behind the above physical ideas, let us neglect for the
moment the $D$-branes and look at the topological strings which are described by $1+1$ TQFT's. Let us recall some
basic notions from algebra. Let $k$ be a commutative ring and $L$ a 
Lie algebra over $k$ endowed with a {\em surgenerate} bilinear form $\b$, i. e. 
a mapping $\b: L \times L \rightarrow k$ which defines an isomorphism
of $k$-modules 
\be
\b(x,y) = \left\langle \psi(x),y \right\rangle
\label{sm},
\ee
where $L^*$ is the dual to $L$, i. e. $L^* = Hom_k(L,k)$. 
\\
\\
{\bf Definition 1.}  $(L,\b)$ is said to be {\em quasi-Frobenius algebra} if
$L$ is finitely generated projective $k$-module and 
\be
\b([x,y],z) + \b([y,z],x) + \b([z,x],y)= 0,~\forall x,y,z, \in L.
\label{liequasifrobenius}
\ee
{\bf Definition 2.} The quasi-Frobenius algebra $(L,\b)$ is said to be 
{\em Frobenius algebra} if there is an element $\phi \in L^*$ such 
that the bilinear form
\be
\b_{\phi} (x,y) = \phi([x,y]),~ \forall x,y \in L
\label{liefrobenius}
\ee
is surgenerate.
\\ 

The mapping $\phi$ is called the {\em Frobenius homomorphism} of $L$.
The definition of Frobenius algebra does not apply only to Lie algebras. In fact, the following more general definition
can be given. 
\\
\\
{\bf Definition 3.} An associative $k$-algebra $A$ is said to be $Frobenius$
if $A$ is finitely generated $k$-module and there exists $\phi \in A^*$
such that the bilinear form $\b_{\phi}(x,y) = \phi(xy),~  \forall x,g \in A$,
where $\phi$ is a Frobenius homomorphism.
\\

The importance of these notions for TQFT is revealed by the following theorem \cite{fq,bb}
\\
\\
{\bf Theorem 1.:} {\em There is a one-to-one correspondence between the set of all 1+1 TQFT's and the finite-dimensional
Frobenius algebras.}
\\
\\
The content of the above theorem can be better understood if we recall the definition of a TQFT.
\\
\\
{\bf Definition 4.} A $d+1$ dimensional TQFT is defined by the following collection of data \cite{bb}:\\
1. To any $d$ dimensional manifold $N$ without boundary is assigned a finite-dimensional vector space $\t(N)$.\\
2. To any $(d+1)$ manifold $M$ (possibly with boundary) is assigned a vector $\t(M)$ in the vector space $\t(\6 M)$.\\
3. To any homeomorphism of $d$ dimensional manifolds $f: N \rightarrow N'$ is assigned an isomorphism of vector spaces
$f_* : \t(N) \rightarrow \t(N')$.\\
4. The following functorial isomorphisms hold
\bea
\t(\bar{N})&\rightarrow& \t(N)^* , \label{1i}\\
\t(\emptyset)&\rightarrow& k , \label{2i}\\
\t(N_1 \sqcup N_2) &\rightarrow & \t(N_1) \otimes \t(N_2), \label{3i}
\eea
where $\bar{N}$ is the manifold $N$ with the opposite orientation. The isomorphisms given by the relations
(\ref{1i}), (\ref{2i}) and (\ref{3i}) are compatible with each other and with the commutativity, associativity and
unit morphism.
The above set of data define a $d+1$ TQFT if they satisfy the axioms of {\bf functoriality}, {\bf gluing},
{\bf normalization} and {\bf border} (see the Appendix.)

The abstract definition of a TQFT formalizes the more intuitive concepts that the spacetime manifold 
interpolates among its boundaries and that the physical fields belong to some spaces defined on the
respective manifolds. The physical content of the theory is contained in the partition function which should
satisfy some tensor properties derived from the additive property of the action and should associate a number to
the spacetime manifold and to any of its boundaries. This number depends only on the topological properties of 
the manifold. Mathematically, this definition assets that TQFT is a category (a bicategory) and Theorem 1
allows us to compute the their properties in terms of the Frobenius algebras.

\subsection{Topological Strings}

One can think of the topological strings as $1+1$ TQFT. However, a full string theory contains products of 
two topologically
distinct sectors: the open string and the closed string sector, respectively. In these sectors the fields are described
by vector spaces on the $2d$ manifold and different vector spaces on the boundary and the orientation of the
surfaces play an important role. To take into account this new data, one should modify the category used in the
previous section and which is appropriate to describe only the closed string sector. To this end, one 
has to define a vector space $V(\S )$ where $\S$ is the two-dimensional world-sheet with
$\6\S_1$ and $\6\S_2$ compact, oriented, one-dimensional boundaries of $\S$ and 
\be
V(\6\S_1 \sqcup \6\S_2) = V(\6\S_1)\otimes V( \6\S_2).
\label{vb}
\ee
In the open string sector the boundaries are closed intervals and the vector space is $V(I)$ while in the
closed string sector the boundaries are circles and the vector space is $V(S^1)$. However, open and closed
strings interact and products of intervals and circles represent possible boundaries. The physical interpretation
of the vector space $V(\6\S)$ is that of the space of states on the boundary. Eventually, we would like that some
of these states live on the $D$-branes.

As we have seen in the previous subsection, there is a Frobenius algebra that dictates the way in which the 
$2d$ manifold be composed such that the $2d$ TQFT be compatible with the geometric gluing  of $2d$ surfaces.
Thus, the algebraic structure obtained in this way results from the 
very definition of the TQFT. Therefore, one should impose some consistency conditions on the state spaces such that
they be compatible with the Frobenius algebra of the world-sheet composition. Let us analyze what are the elements of
this construction. Firstly, we have the pure open string sector and the closed string sector. In this sectors the world-sheets
are two-dimensional manifolds which interpolate among boundaries of just one type, i. e. 
mappings like 
\be
\S_{\mbox{\small open}} : I \rightarrow I~~,~~ \S_{\mbox{\small closed}}: S^1 \rightarrow S^1.
\label{oocc}
\ee
Beside these sectors the TQFT contains open-closed and closed-open sectors and the world-sheets in these sectors are 
cobordisms among boundaries which are disjoint unions of open and closed string boundaries, respectively,
\be
\6\S = \left(\sqcup_{m}(I)^m \right)\sqcup \left( \sqcup_{n}(S^1)^n \label{ocb}\right).
\ee
The basic scattering processes in the open-closed and closed-open sectors are given by mappings like
\be
\S_{\mbox\small open-closed}: I \rightarrow S^1 ~~,~~ \S_{\mbox{\small closed-open}}: S^1 \rightarrow I.
\label{occo}
\ee
The following relation between the above mappings hold 
\be
\bar{\S}_{\mbox{open-closed}}=\S_{\mbox{closed-open}}~~,~~
\bar{\S} \circ \S = 1.
\label{relocco}
\ee
The relations (\ref{oocc}), (\ref{occo}) and (\ref{relocco}) show that the state spaces satisfy a certain algebra. 
In order find what this algebra is, note that $V(I)$ and $V(S^1)$ are commutative and that there is a neuter element
with respect to multiplication (unit). If we take now $V(S^1)$, there are two remarkable maps from and to the field of 
complex numbers, respectively,
\be
E_c: \CC \rightarrow V(S^1)~~,~~ \b_c : V(S^1) \rightarrow \CC .
\ee
The interpretation of $E_c$ is that it defines the unit element for the algebra of state spaces.
 The map $\b_c$ is a
non-degenerate bilinear trace. The properties of the two maps can be verified if they are
represented as the cobordisms generated by the disk 
\be
\b_c::~~ B^2: S^1 \rightarrow \emptyset ~~~~,~~~~ E_c::~~ B^2 : \emptyset \rightarrow S^1.
\label{buc}
\ee
From these properties we recognize the structure of the Frobenius algebra. The bilinear
trace $\b_o$ for the state space $V(I)$ is given by the following relation \cite{gs}
\be
\b_o = \b_c \circ i^* ~~,~~ i: V(S^1) \rightarrow V(I)~~,~~i^*: V(I) \rightarrow V(S^1).
\label{ouc}
\ee
Here, $i^*$ denotes the homomorphism of modules adjoint to the map $i$. If one considers the open 
strings, one can prove the following \cite{gs}
\\
\\
{\bf Theorem 2.} {\em A two-dimensional open topological string theory is given by the following
set of data:\\
1. A commutative Frobenius algebra $(V(S^1),\b_c)$.\\
2. A commutative Frobenius algebra $(V(I),\b_o)$.\\
3. A map $i: V(S^1) \rightarrow V(I)$ such that $i(1)= 1$ and $Im(i) \subset C(V(I)) $ where $C(V(I))$ is
the center of $V(I)$, i. e. all commuting elements of $V(I)$.}
\\

The result of computations in a topological open string theory is usually a number associated to all
surfaces topologically equivalent with any of the surfaces from the loop-expansion of the partition function defined in
(\ref{pf}) for opens string. From the physical point of view, these numbers correspond to the elements of the $S$-matrix 
that describes scattering processes among arbitrary number of strings on a determined world-sheet and in a fixed 
spacetime background. The open strings that enter in the process are described by boundaries of $I$ type. There
are also closed string described by boundaries of the $S^1$ type.
Thus, the elements of the $S$-matrix depend on
the number of circles and the genus $g$ of the world-sheet.

Let us turn our attention to the full open and closed string theory. As we have seen in the previous subsection,
the closed strings can be described by a TQFT which, at its turn, can be interpreted either as a category or as a 
commutative Frobenius algebra. Therefore, the open string theory as stated in the Theorem 2 should be supplemented 
with the Frobenius algebra from the closed string sector and with the compatibility conditions for 
all sectors. Summarizing these observations,
let us state the theorem that defines a open and closed topological string theory \cite{gm}
\\
\\
{\bf Theorem 3} {\em A two-dimensional (oriented) open and closed topological string theory is given by the 
following data:\\
1. A commutative Frobenius algebra $A$ (for the TQFT.)\\
2. A commutative Frobenius algebra $V(\6\S)$ for each boundary condition.\\
3. A homomorphism $i_{V(\6\S)}: \CC \rightarrow C(V(\6\S))$ where $C(V(\6\S))$ is the center of $V(\6\S)$ and 
$i_{V(\6\S)}=1$.\\
4. For any two algebras $V(\6\S_1)$ and $V(\6\S_2)$ the corresponding morphism is given by the composition
\be
\pi^{1}_{2}:V(\6\S_1)\rightarrow V(\6\S_2)~~,~~
\pi^{1}_{2} = i_{V(\6\S_2)} \circ i_{V(\6\S_1)}^* .
\label{mapalg}
\ee }

The known examples of open and closed string theories have a semi-simple Frobenius algebra $A$.
Important examples are represented by simplified versions of topological strings on Calaby-Yau
manifolds, useful for understanding the compactification down to four dimensions (see \cite{psa,mrd1,mrd2,gs}
and references therein.)

Theorem 3 and the known examples show that there is an algebraic elementary structure behind topological
string theory. Physical properties of such theories are encoded in certain algebraic objects as K-theory groups.

\section{Topological Field String Theories and $D$-branes}

In the perturbative limit of string theory, the $D$-branes are represented by a set of Neumann and Dirichlet 
boundary conditions which break the Lorentz invariance of spacetime but preserve the conformal invariance
of the world-sheet theory. Since the flat spacetime theory is the guiding principle in generalizing the 
string theory, it is postulated that the $D$-branes in any CFT in two dimensions are defined by the same 
property. Namely, they are arbitrary local boundary conditions of the two-dimensional CFT that preserve the conformal
symmetry on the boundary. A natural technique to extract information from such concept is the renormalization group 
(RG) when applied to the space of all two-dimensional theories which are conformally invariant in the bulk but may 
have the conformal invariance broken on the boundary. This point of view is extensively presented in \cite{gm}
and we refer the reader to this reference for details. Instead, let me speculate on a different line of thought on
which we would like to generalize the definitions of topological strings and $D$-branes. 

Let us systematize the ideas from the previous section. A two-dimensional topological string theory is a choice
of two Frobenius algebras $A_1$ and $A_2$ for the TQFT and the other one for the set of state spaces associated to 
boundary conditions, respectively,
and a map
${\cal{I}}= \{ i_{V(\6\S)} \}$ (the set of all maps $i$ for all boundaries)
with adjoint. These algebras are related to the modular category of vector spaces over one-dimensional
compact boundaries (freely constructed from unions of elements of the {\em basis} $\{ I , S^1 \}$) with $2d$ surfaces
(morphisms) among them. The $D$-branes are boundary conditions that preserve the conformal invariance and, therefore,
induce a subset 
\be
\DD(\6\S) \hookrightarrow \VV (\6\S)
\label{dbc}
\ee
of the set of all state spaces for a given boundary $\6\S$.
This implies that the $D$-branes select a subcategory from the original modular category in which the 
objects are consistent with the conformal invariance. Next, note that since the composition of surfaces can
hold at left and right and the same hold for states in the state spaces, the category of open and closed strings 
$\SS$ is actually a bicategory. Furthermore, defining a tensor product in a natural way makes $\SS$ a tensor category,
too, and since the tensor product is associative the category is also strict (see the Apendix.) 
Now let us recall the following result
\cite{pdf}
\\
{\bf Theorem 4} { \em Let $\AA$ be a (strict) tensor category and $A$ be a Frobenius algebra in $\AA$. Then there exists
and almost bicategory $\O$ such that:\\
1. $\mbox{Obj}(\O) = \{ \G , \L \}$ where $\G, \L$ are 2- or 1-morphism, respectively, 
in the 2-category objects from \AA.\\
2. There is an isomorphism $I: \AA \rightarrow \mbox{Hom}_{\O}(\G ,\G )$ of tensor categories.\\
3. There is an 1-morphism $J: \L \rightarrow \G$ and $\bar{J}: \G \rightarrow \L$ such the $J\bar{J} = \cal{I}(A)$.}
\\

The almost bicategory $\O$ can be turned into a bicategory if a further requirement is imposed 
on the Frobenius algebra to produce the lacking unit morphism $\L \rightarrow \L$.

Now let us apply the Theorem 4 to the topological strings. According to it, one can construct two almost bicategories 
$\O_1$ and $\O_2$. The objects in this category are morphisms inside the objects in the tensor categories. At
their turn, these objects are $2d$ TQFT and spaces of states, respectively. Therefore, it is natural to think of 
morphisms between TQFT and between spaces of states which are associated to open and closed topological strings 
as belonging to the category of topological string field theories. 
In particular, from the Theorem 4 we have the following identifications for each Frobenius algebra
\bea
\mbox{Hom}_{\O}(\G , \G) &=& \mbox{Obj}(\AA)\label{sft1},\\
\mbox{Hom}_{\O}(\L , \G) &=& \{ XJ,~ X \in \mbox{Obj}(\AA) \}\label{sft2},\\
\mbox{Hom}_{\O}(\G , \L) &=& \{ JX,~ X \in \mbox{Obj}(\AA) \}\label{sft3},\\
\mbox{Hom}_{\O}(\L , \L) &=& \{ \bar{J}X J, ~X\in \mbox{Obj}(\AA)\} \label{sft4}.
\eea
From the first relation (\ref{sft1}) we see that if the objects of the new category are maps between 
$2d$ topological open and closed strings (topological string fields) then the corresponding morphisms are 
TQFT, too. It would be interesting to investigate what is the exact interpretation of the lift of the 
$\II$ map in the category $\O$ and the implication of the structure 
$V(\6\S_1 \sqcup \6\S_2) \sim \mbox{Mor}(\6\S_1,\6\S_2)$ which is an additive category \cite{gs}, since it is 
known that the monoidal Morita equivalence implies equivalent quantum doubles and the same state sum of closed
oriented three-manifolds \cite{pdf}.

Let us briefly discuss the implications of the above speculations on the $D$-branes. From the relation (\ref{dbc})
we see that introducing $D$-branes in a given topological string theory actually restricts the Frobenius algebra
associated with the state spaces to certain closed set of spaces compatible with the conformal symmetry. The
exact content of this set is not know, but by using the renormalization group methods one can identify, in principle,
the points in the moduli space where the conformal symmetry is unbroken. By the construction given in the 
Theorem 4, the $D$-branes are lifted to a subset of the almost bicategory $\O_2$ which has as objects the morphisms
among the spaces of states consistent with the conformal symmetry. In particular, one could interpret these morphisms
as $D$-branes in the topological string field theory. The maps among the $D$-branes are, according with the relation
(\ref{sft1}), objects from the tensor category of state spaces. It is of interest to see if there is any Morita 
equivalence inherited by $D$-branes from the monoidal equivalence mentioned above. We hope to report in more detail
on the above ideas elsewhere.

{\bf Acknowledgments:}
I would like to thank to the organizers of the {\em Second International Conference on 
Fundamental Interactions} for invitation. I also acknowledge to J. A. Helayel-Neto and 
S. Alves for hospitality at DCP-CBPF where part of this paper was done and to E. L. Gra\c{c}a for
discussions.
This work is supported by FAPESP Grant 02/05327-3.

\section{ Appendix}

In this appendix we are going to give basic mathematical definitions which have not been included in the text. See for
more details \cite{sml}.
\\
{\bf Functoriality Axiom.} If $f: M \rightarrow M'$ is a homeomorphism of $d+1$ dimensional manifolds then
$(f|_{\6 M})_*(\t(M))=\t(M')$.
\\
{\bf Gluing Axiom.} If $M$ is a $d+1$ dimensional manifold, $\6 M = N_1 \sqcup N_2 \sqcup N_3$ and 
$f: N_1 \rightarrow N_2$ is a homeomorphism, then $\t(M') = \psi(\t(M))$,  where $\psi$ is the map
$\psi : \t(N_1) \otimes \t(N_2) \otimes  \t(N_3) \rightarrow \t(N_2)^* \otimes \t(N_2) \otimes \t(N_3) 
\rightarrow \t(N_3)$ and $M' = M/f$ is a $d+1$ dimensional manifold obtained from $M$ by identifying 
$N_1$ with $\bar{N}_2$ usinf $f$, i. e. by gluing $N_1$ to $N_2$.
\\
{\bf Normalization Axiom.} Let $I$ be an interval and $N$ be a $d$ dimensional manifold. Then $\6(I \times N) = N \sqcup 
\bar{N}$ and we require that $\t(I \times N)$ equals the image of $id_{\t(N)}$ in $\t(\bar{N}) \otimes \t(N) \otimes
\t(N_3) \rightarrow \t(N_3)$.
\\
{\bf Border Axiom.} If $B^{d+1}$ is the unit ball in $\RR^{d+1}$ and $S^d = \6 B^{d+1}$ then $\t(S^d)=k$ and 
$\t(B^{d+1}) = 1 \in k$.
\\
{\bf Strict Tensor Category} A tensor category is said to be {\em strict} if the tensor product $\otimes$ is
associative, i. e. $(X\otimes Y)\otimes Z = X \otimes (Y \otimes Z)$.

\end{document}